\documentstyle[twoside,psfig]{article}

\def\beq{\begin{equation}}
\def\eeq{\end{equation}}
\def\beqa{\begin{eqnarray}}
\def\eeqa{\end{eqnarray}}

\catcode`\@=11
\long\def\@makefntext#1{
\protect\noindent \hbox to 3.2pt {\hskip-.9pt  
$^{{\eightrm\@thefnmark}}$\hfil}#1\hfill}		

\def\@makefnmark{\hbox to 0pt{$^{\@thefnmark}$\hss}}	
	
\def\ps@myheadings{\let\@mkboth\@gobbletwo
\def\@oddhead{\hbox{}
\rightmark\hfil\eightrm\thepage}   
\def\@oddfoot{}\def\@evenhead{\eightrm\thepage\hfil
\leftmark\hbox{}}\def\@evenfoot{}
\def\sectionmark##1{}\def\subsectionmark##1{}}

\oddsidemargin=\evensidemargin
\newcounter{sectionc}\newcounter{subsectionc}\newcounter{subsubsectionc}
\renewcommand{\section}[1] {\vspace{12pt}\addtocounter{sectionc}{1} 
\setcounter{subsectionc}{0}\setcounter{subsubsectionc}{0}\noindent 
	{\tenbf\thesectionc. #1}\par\vspace{5pt}}
\renewcommand{\subsection}[1] {\vspace{12pt}\addtocounter{subsectionc}{1} 
	\setcounter{subsubsectionc}{0}\noindent 
	{\bf\thesectionc.\thesubsectionc. {\kern1pt \bfit #1}}\par\vspace{5pt}}
\renewcommand{\subsubsection}[1] {\vspace{12pt}\addtocounter{subsubsectionc}{1}
	\noindent{\tenrm\thesectionc.\thesubsectionc.\thesubsubsectionc.
	{\kern1pt \tenit #1}}\par\vspace{5pt}}
\newcommand{\nonumsection}[1] {\vspace{12pt}\noindent{\tenbf #1}
	\par\vspace{5pt}}

\newcounter{appendixc}
\newcounter{subappendixc}[appendixc]
\newcounter{subsubappendixc}[subappendixc]
\renewcommand{\thesubappendixc}{\Alph{appendixc}.\arabic{subappendixc}}
\renewcommand{\thesubsubappendixc}
	{\Alph{appendixc}.\arabic{subappendixc}.\arabic{subsubappendixc}}

\renewcommand{\appendix}[1] {\vspace{12pt}
        \refstepcounter{appendixc}
        \setcounter{figure}{0}
        \setcounter{table}{0}
        \setcounter{lemma}{0}
        \setcounter{theorem}{0}
        \setcounter{corollary}{0}
        \setcounter{definition}{0}
        \setcounter{equation}{0}
        \renewcommand{\thefigure}{\Alph{appendixc}.\arabic{figure}}
        \renewcommand{\thetable}{\Alph{appendixc}.\arabic{table}}
        \renewcommand{\theappendixc}{\Alph{appendixc}}
        \renewcommand{\thelemma}{\Alph{appendixc}.\arabic{lemma}}
        \renewcommand{\thetheorem}{\Alph{appendixc}.\arabic{theorem}}
        \renewcommand{\thedefinition}{\Alph{appendixc}.\arabic{definition}}
        \renewcommand{\thecorollary}{\Alph{appendixc}.\arabic{corollary}}
        \renewcommand{\theequation}{\Alph{appendixc}.\arabic{equation}}
        \noindent{\tenbf Appendix \theappendixc #1}\par\vspace{5pt}}
\newcommand{\subappendix}[1] {\vspace{12pt}
        \refstepcounter{subappendixc}
        \noindent{\bf Appendix \thesubappendixc. {\kern1pt \bfit #1}}
	\par\vspace{5pt}}
\newcommand{\subsubappendix}[1] {\vspace{12pt}
        \refstepcounter{subsubappendixc}
        \noindent{\rm Appendix \thesubsubappendixc. {\kern1pt \tenit #1}}
	\par\vspace{5pt}}

\newcommand{\textlineskip}{\baselineskip=13pt}
\newcommand{\smalllineskip}{\baselineskip=10pt}

\def\eightcirc{
\begin{picture}(0,0)
\put(4.4,1.8){\circle{6.5}}
\end{picture}}
\def\eightcopyright{\eightcirc\kern2.7pt\hbox{\eightrm c}}

\def\abstracts#1#2#3{{
	\centering{\begin{minipage}{4.5in}\baselineskip=10pt\footnotesize
	\parindent=0pt #1\par 
	\parindent=15pt #2\par
	\parindent=15pt #3
	\end{minipage}}\par}} 

\renewenvironment{thebibliography}[1]
	{\frenchspacing
	 \ninerm\baselineskip=11pt
	 \begin{list}{\arabic{enumi}.}
	{\usecounter{enumi}\setlength{\parsep}{0pt}
	 \setlength{\leftmargin 12.7pt}{\rightmargin 0pt} 
	 \setlength{\itemsep}{0pt} \settowidth
	{\labelwidth}{#1.}\sloppy}}{\end{list}}

\newcounter{itemlistc}
\newcounter{romanlistc}
\newcounter{alphlistc}
\newcounter{arabiclistc}

\newcommand{\fcaption}[1]{
        \refstepcounter{figure}
        \setbox\@tempboxa = \hbox{\footnotesize Fig.~\thefigure. #1}
        \ifdim \wd\@tempboxa > 5in
           {\begin{center}
        \parbox{5in}{\footnotesize\smalllineskip Fig.~\thefigure. #1}
            \end{center}}
        \else
             {\begin{center}
             {\footnotesize Fig.~\thefigure. #1}
              \end{center}}
        \fi}

\newcommand{\tcaption}[1]{
        \refstepcounter{table}
        \setbox\@tempboxa = \hbox{\footnotesize Table~\thetable. #1}
        \ifdim \wd\@tempboxa > 5in
           {\begin{center}
        \parbox{5in}{\footnotesize\smalllineskip Table~\thetable. #1}
            \end{center}}
        \else
             {\begin{center}
             {\footnotesize Table~\thetable. #1}
              \end{center}}
        \fi}

\def\@citex[#1]#2{\if@filesw\immediate\write\@auxout
	{\string\citation{#2}}\fi
\def\@citea{}\@cite{\@for\@citeb:=#2\do
	{\@citea\def\@citea{,}\@ifundefined
	{b@\@citeb}{{\bf ?}\@warning
	{Citation `\@citeb' on page \thepage \space undefined}}
	{\csname b@\@citeb\endcsname}}}{#1}}

\newif\if@cghi
\def\cite{\@cghitrue\@ifnextchar [{\@tempswatrue
	\@citex}{\@tempswafalse\@citex[]}}
\def\citelow{\@cghifalse\@ifnextchar [{\@tempswatrue
	\@citex}{\@tempswafalse\@citex[]}}
\def\@cite#1#2{{$\null^{#1}$\if@tempswa\typeout
	{IJCGA warning: optional citation argument 
	ignored: `#2'} \fi}}

\def\pmb#1{\setbox0=\hbox{#1}
	\kern-.025em\copy0\kern-\wd0
	\kern.05em\copy0\kern-\wd0
	\kern-.025em\raise.0433em\box0}

\def\fnt#1#2{\footnotetext{\kern-.3em
	{$^{\mbox{\scriptsize #1}}$}{#2}}}

\def\fpage#1{\begingroup
\voffset=.3in
\thispagestyle{empty}\begin{table}[b]\centerline{\footnotesize #1}
	\end{table}\endgroup}

\font\tenrm=cmr10
\font\tenit=cmti10 
\font\tenbf=cmbx10
\font\bfit=cmbxti10 at 10pt
\font\ninerm=cmr9

\font\eightrm=cmr8

\def\qed{\hbox{${\vcenter{\vbox{			
   \hrule height 0.4pt\hbox{\vrule width 0.4pt height 6pt
   \kern5pt\vrule width 0.4pt}\hrule height 0.4pt}}}$}}

\begin{document}


\normalsize\textlineskip
\thispagestyle{empty}
\setcounter{page}{1}


\begin {flushright}
FSU-HEP-20000901\\
\end {flushright} 

\vspace*{0.5truein}

\fpage{1}
\centerline{\bf TOP QUARK TOTAL AND DIFFERENTIAL CROSS SECTIONS}
\vspace*{0.035truein}
\centerline{\bf AT NNLO AND NNLL\footnote{Presented at DPF2000, Columbus, 
Ohio, August 9-12, 2000; to appear in the Proceedings.}}
\vspace*{0.37truein}
\centerline{\footnotesize NIKOLAOS KIDONAKIS\footnote{This work was supported 
in part by the US Department of Energy.}}
\vspace*{0.015truein}
\centerline{\footnotesize\it Department of Physics, Florida State University,}
\baselineskip=10pt
\centerline{\footnotesize\it Tallahassee, FL 32306-4350, USA}
\vspace*{10pt}
\vspace*{0.21truein}
\abstracts{I present recent NNLO-NNLL results for top quark hadroproduction at
the Tevatron. The total cross section as well as transverse
momentum and rapidity distributions are shown. } {} {}

\vspace*{1pt}\textlineskip	
\section{Introduction}	
\vspace*{-0.5pt}
\noindent

The top quark production cross section at the Tevatron receives
contributions from the threshold region where there are 
potentially large corrections. Near threshold for the production of 
the $ t {\bar t}$ pair, there is  restricted phase space
for real gluon emission. This results in an
incomplete cancellation of infrared divergences 
between real and virtual graphs which manifests itself in large 
logarithmic corrections.

At $n$th order in $\alpha_s$, these corrections are of the form
$[(\ln^k(s_4/m^2))/s_4]_+$, $k \le 2n-1$, with $m$ the top mass and
$s_4 \equiv s+t_1+u_1$, $ s_4 \rightarrow 0$ 
at threshold. They can be formally resummed in moment space
at next-to-leading logarithmic (NLL) or higher accuracy to all orders 
in perturbation theory,$^{1-7}$
but a prescription is required in practice to obtain numerical results.
In this talk, I present the next-to-next-to-leading order (NNLO) threshold 
corrections for $t {\bar t}$ production at next-to-next-to-leading logarithmic
(NNLL) accuracy.$^{8}$ These follow from the two-loop 
expansion of the resummed cross section without need of a prescription.

The factorized top quark production cross section can be written as
a convolution of parton distributions $\phi$ with the perturbative
cross section $\hat{\sigma}$: 
$\sigma_{h_1h_2\rightarrow t{\bar t}}
=\sum_f
\phi_{f_i/h_1} \otimes \phi_{f_j/h_2}
\otimes\hat{\sigma}_{f_i f_j\rightarrow t{\bar t}}\, .$
The main partonic processes involved are
$q {\bar q} \rightarrow t {\bar t}$ and $gg \rightarrow t {\bar t}$.
By taking moments, the above expression for the cross section simplifies to 
${\tilde{\sigma}}_{f_i f_j \rightarrow t{\bar t}}(N)
={\tilde{\phi}}_{f_i/f_i}(N)\,  {\tilde{\phi}}_{f_j/f_j}(N)\,
\hat{\sigma}_{f_i f_j \rightarrow t{\bar t}}(N)$
where the moments are defined by 
$\hat{\sigma}(N)=\int (ds_4/s) \, 
e^{-Ns_4/s} {\hat\sigma}(s_4)$, etc., with $N$ the moment variable.
Under moments,
$[(\ln^{2n-1}(s_4/m^2))/s_4]_+ \rightarrow \ln^{2n}N$, and our
goal becomes to resum logarithms of $N$.

We may refactorize the cross section$^{1,3,5}$ as
${\tilde{\sigma}}_{f_i f_j \rightarrow t{\bar t}}(N)
={\tilde{\psi}}_{f_i/f_i}(N)\,  {\tilde{\psi}}_{f_j/f_j}(N)\,
\newline \times H_{IJ} \;  {\tilde{S}}_{JI}(m/(N\mu_F))$, where the
$\psi$ are center-of-mass parton distributions,
$H$ is the hard-scattering matrix (N-independent),
and $S$ is the soft-gluon function that describes noncollinear
soft gluon emission. $H$ and $S$ are matrices in the space of color 
exchanges and depend on the partonic process.
 
Solving for the perturbative cross section ${\hat{\sigma}}$, we have
${\hat{\sigma}}(N)=({\tilde{\psi}}/{\tilde{\phi}})^2 \; {\rm Tr}\,
[H{\tilde{S}}]$.
After resumming the N-dependence in $\psi/\phi$ and $S$, we obtain the
resummed top quark cross section in moment space at NLL accuracy:
\beqa
\tilde{{\sigma}}_{f_i f_j \rightarrow t{\bar t}}(N) &=&   
\exp\left[ E^{(f_i)}(N_i,\mu_F)+E^{(f_j)}(N_j,\mu_F)\right] 
\exp\left[4\int_{\mu_R}^{m}\frac{d\mu'}{\mu'} 
\beta(\alpha_s(\mu'^2))\right]
\nonumber\\ && \hspace{-30mm} 
\times \exp \left[2\int_{\mu_F}^{m} {d\mu' \over \mu'}
\left(\gamma_i(\alpha_s(\mu'^2))
+\gamma_j(\alpha_s(\mu'^2))\right)\right] \; {\rm Tr} \left \{
H\left(\alpha_s(\mu_R^2)\right) \; \right.
\nonumber\\ && \hspace{-30mm} \times \; \left.
\bar{P} \exp \left[\int_m^{m/N} {d\mu' \over \mu'} \;
\Gamma_S^\dagger\left(\alpha_s(\mu'^2)\right)\right] \;
{\tilde S}(1) \;
P \exp \left[\int_m^{m/N} {d\mu' \over \mu'}\; \Gamma_S
\left(\alpha_s(\mu'^2)\right)\right] \right\}\, ,
\eeqa
where the incoming parton $N$-dependence is in $E^{(f)}$,
$\gamma_i$ is the anomalous dimension of the field $\psi_i$,
and $\Gamma_S$ is the process-dependent soft anomalous dimension 
matrix which has been calculated explicitly at one-loop.$^{1}$

The resummed cross section is then expanded to  
NNLO$^{8}$ and even higher orders,$^{9}$
thus avoiding prescription dependence and unphysical terms.$^{9}$
After matching with the exact NLO cross section$^{10}$ one can determine the
next-to-next-to-leading logarithms and thus obtain the 
NNLO-NNLL corrections to top quark production.
Similar expansions have already been presented for electroweak-boson,$^{11}$
direct photon,$^{12}$ and jet$^{13}$ hadroproduction.

\section{NNLO-NNLL corrections}
\noindent

We now expand the resummed cross section in 
the $\overline {\rm MS}$ scheme to two-loops in single-particle 
inclusive kinematics.$^{8}$ 
The NLO threshold corrections agree with exact NLO results.$^{10}$

The NNLO-NNLL threshold corrections in the $q {\bar q}$ channel are
\beqa
&&\hspace{-6mm}
{\hat \sigma}^{\overline {\rm MS} \, (2)}_{q{\bar q}\rightarrow t{\bar t}}
(s_4,m^2,t_1,u_1)=\sigma^B_{q{\bar q}\rightarrow t{\bar t}} 
\left(\frac{\alpha_s(\mu_R^2)}{\pi}\right)^2 
\left\{8 C_F^2 \left[\frac{\ln^3(s_4/m^2)}{s_4}\right]_{+} \right.
\nonumber \\ && \hspace{-9mm}
{}+\left[\frac{\ln^2(s_4/m^2)}{s_4}\right]_{+} C_F \left[-\beta_0 
+12 \left({\rm Re}\Gamma_{22}'-C_F+C_F\ln\left(\frac{sm^2}{t_1u_1}\right)
-C_F\ln\left(\frac{\mu_F^2}{m^2}\right)\right)\right]
\nonumber \\ &&  \hspace{-9mm}
{}+\left[\frac{\ln(s_4/m^2)}{s_4}\right]_{+}
\left[4\left[{\rm Re} \Gamma'_{22}- C_F
-C_F \ln\left(\frac{t_1u_1}{sm^2}\right)
-C_F \ln\left(\frac{\mu_F^2}{m^2}\right)\right]^2 \right.
\nonumber \\ && \hspace{-9mm}
{}+4\Gamma'_{12}\Gamma'_{21}
-\beta_0\left[{\rm Re} \Gamma'_{22}-C_F 
-C_F \ln\left(\frac{t_1 u_1}{sm^2}\right)
-C_F\ln\left(\frac{\mu_R^2}{m^2}\right)\right]
\nonumber \\ && \hspace{-9mm}\left. \left.
{}+2C_F K -16 \zeta_2 C_F^2 \right] \right\}
+4 C_F \frac{\alpha_s(\mu_R^2)}{\pi}
\, \sigma^{(1) \, q {\bar q} \, {\rm S+V}}_{{\overline {\rm MS}}\,
{\rm exact}} \left[\frac{\ln(s_4/m^2)}{s_4}\right]_{+}
+{\cal{O}}\left(\left[\frac{1}{s_4}\right]_+\right) \, ,
\eeqa
where $\sigma^{(1) \, q {\bar q} \, {\rm S+V}}$ denotes the soft
plus virtual $\delta(s_4)$ terms in the NLO cross section.$^{10}$ Note that  
${\hat \sigma}^{\overline {\rm MS} \, (2)}$ actually stands for 
any relevant double differential cross section, such as
$d^2\sigma/(dt_1 du_1)$ or $d^2\sigma/(dp_t^2 dy)$, with appropriate
expressions for the Born term $\sigma^B$.
Analogous results have been derived for the $gg$ channel.$^{8}$

\begin{figure}
\centerline{
\psfig{file=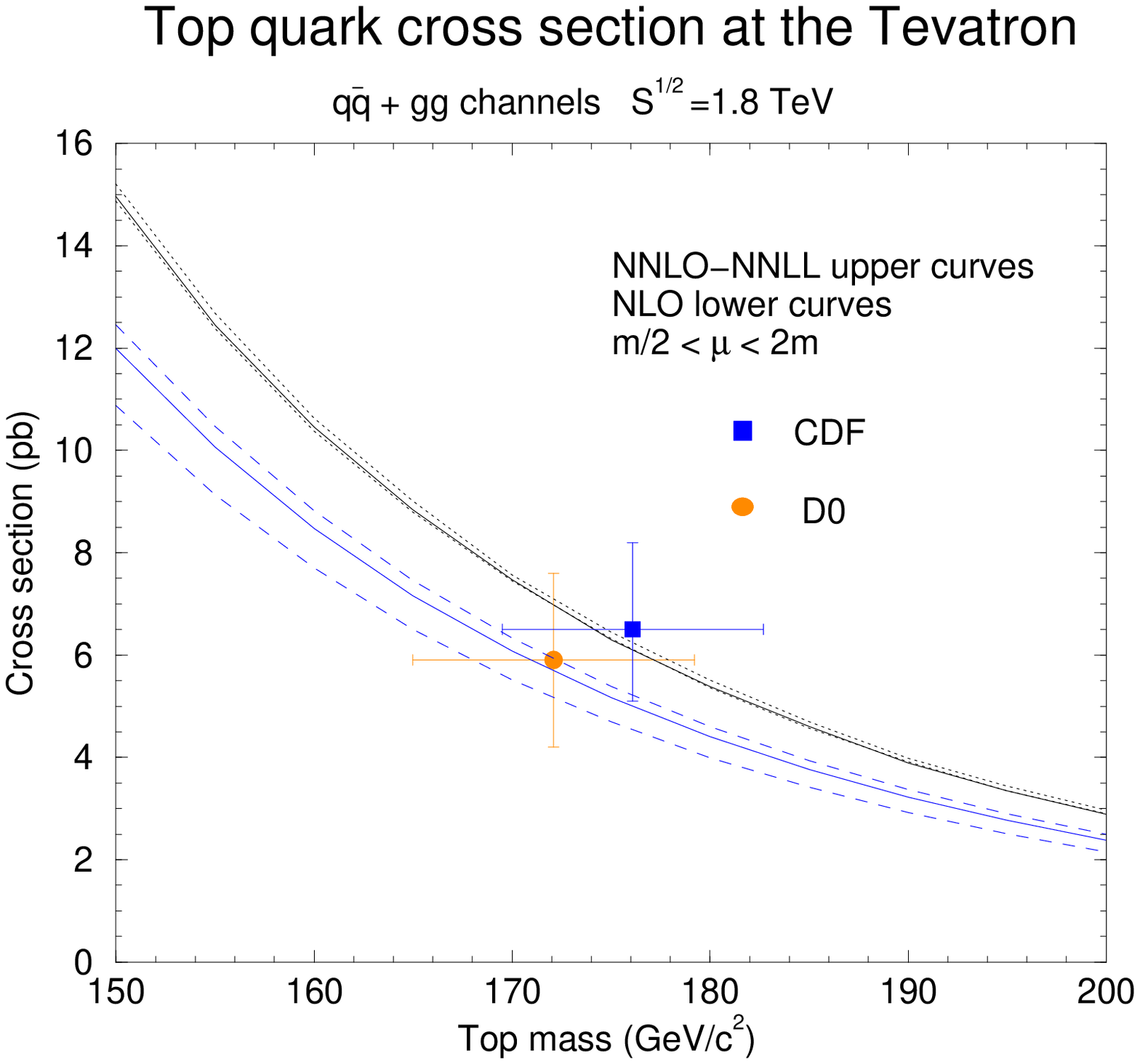,height=3.8in,width=4.5in,clip=}}
{Fig. 1.  Top quark production at the Tevatron:
total cross section.}
\label{fig1}
\end{figure}

In Fig. 1 we show the top quark cross section at the Tevatron
as a function of its mass. We use the CTEQ5M$^{14}$ parton densities.
We note a dramatic decrease of the scale 
dependence when we include the NNLO-NNLL corrections.
The NNLO-NNLL cross section is 6.3 pb versus 5.2 pb at NLO,
an enhancement of over $20 \%$, at $\mu=m$. 
Good agreement is observed with recent results from CDF and D0.
At Run II, with $\sqrt{S}=2.0$ TeV, we predict a NNLO-NNLL
cross section of 8.8 pb versus 7.1 at NLO with $\mu=m$.

\begin{figure}
\centerline{
\psfig{file=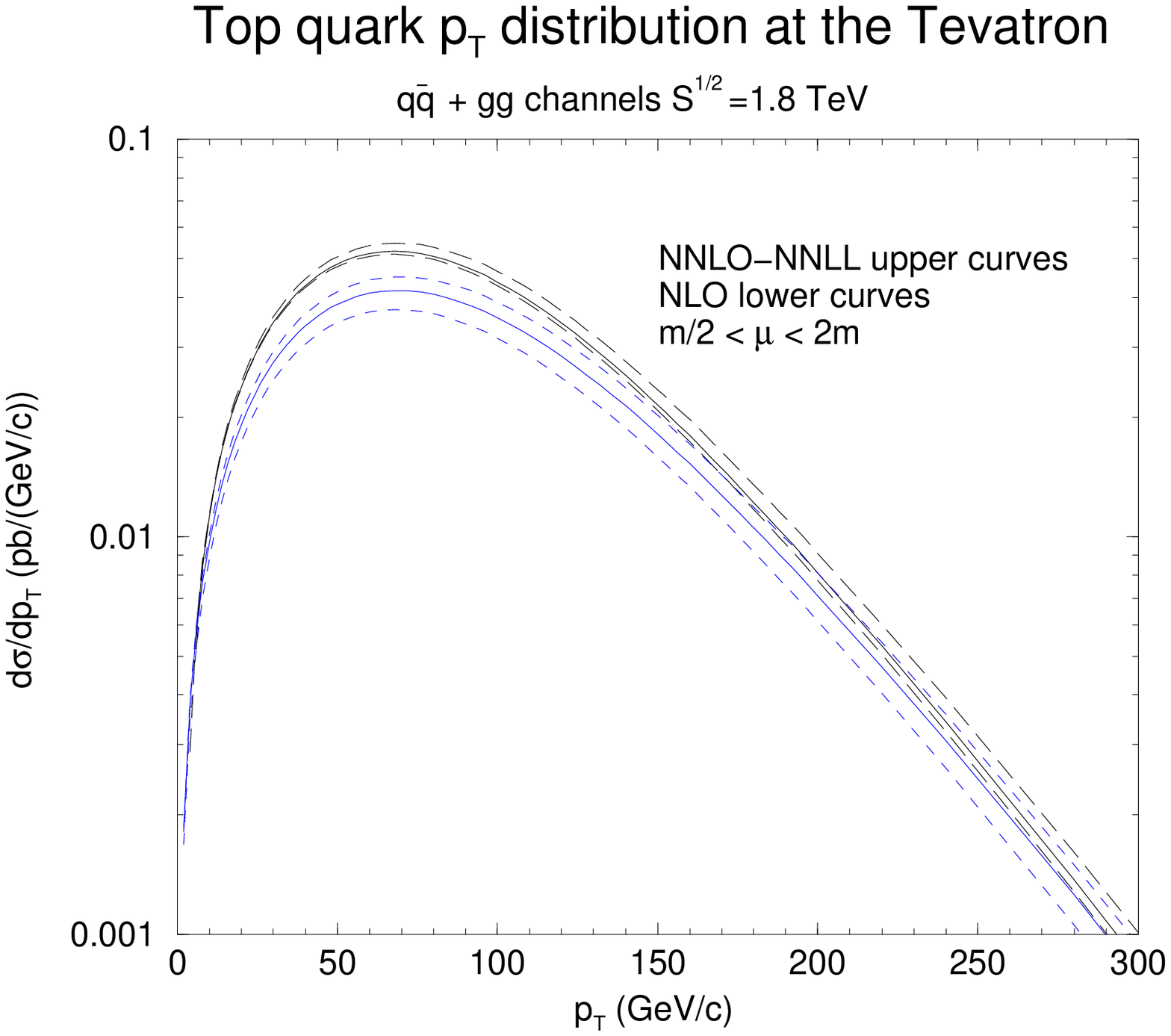,height=3.5in,width=4.5in,clip=}}
{Fig. 2.  Top quark production at the Tevatron:
transverse momentum distribution.}
\label{fig2}
\end{figure}

The top quark transverse momentum distribution is shown in Fig. 2.
We note an overall enhancement at NNLO with little change of shape. 
Similar conclusions are also reached for the rapidity distribution.

\nonumsection{References}

\end{document}